\documentclass[reprint,twocolumn,superscriptaddress,showkeys,
nofootinbib,notitlepage,amsmath,amssymb,floatfix,aps,prd]{revtex4-2}

\usepackage{latexsym,amsmath,amssymb,lmodern,url}
\usepackage{natbib}
\usepackage{color}
\usepackage{multirow}
\usepackage{bm}
\usepackage{array}
\usepackage[normalem]{ulem}
\usepackage{cleveref}
\usepackage{xcolor}

\usepackage{tikz}
\usetikzlibrary{datavisualization}
\usetikzlibrary{datavisualization.formats.functions}
\usetikzlibrary{plotmarks}

\usepackage{mathtools}
\def\de{\delta^{\vphantom{1}}}
\def\bde{{\bar\delta}}

\def\h3{{\displaystyle{\frac 3 2}}}

\newcommand{\bbar}{\overline}

\begin{document}
\title{Diabatic Dynamical Diquark Model of Hidden-Strangeness Tetraquarks}
\author{Shahriyar Jafarzade}
\email{sjafarz2@asu.edu}
\author{Richard F. Lebed}
\email{Richard.Lebed@asu.edu}
\affiliation{Department of Physics, Arizona State University, Tempe,
AZ 85287, USA}
\date{October, 2025}

\begin{abstract}
We generalize our recent analysis of hidden-strangeness tetraquarks within the dynamical diquark model from its adiabatic form (in which each state is described solely by a diquark-antidiquark potential) to its diabatic form (which incorporates effects of di-hadron thresholds upon the states).  We tabulate all relevant thresholds and compute the di-hadron content of each predicted state.  Our results produce no particular hidden-strange tetraquark candidate whose structure is dominated by di-hadron structure, in contrast to the charm sector, where many exotic states are strongly associated with such thresholds : The hidden-strange states tend to remain compact and less influenced by di-hadron thresholds.   Multiple states above 2~GeV with peculiar decay properties, including $\phi(2170)$, $f_2(2340)$, and $\eta(2370)$, continue to serve as excellent hidden-strange tetraquark candidates.
\end{abstract}

\maketitle

\section{Introduction}

In the past two decades, the spectrum of known hadrons has expanded beyond conventional $q\bar q$ mesons and $qqq$ baryons through the observation of numerous \emph{exotic hadrons}.  These novel states include tetraquark and pentaquark candidates in the heavy-quark sector, and have been reported by multiple experimental collaborations.  To note a few specific milestones, Belle discovered $X(3872)$ [now $\chi_{c1}(3872)$] in $B$-meson decays~\cite{Belle:2003nnu}, BESIII observed the charged $Z_c(3900)$ [now $T_{c\bar c 1} (3900)$] in $e^+e^-$ collisions~\cite{BESIII:2013ris}, and LHCb identified hidden-charm pentaquarks such as $P_c(4457)^+$ in $\Lambda_b$ decays~\cite{Aaij:2019vzc}.

Theoretical interpretations for the structure of these hadrons include \emph{compact multiquark\/} states, \emph{hadronic molecules}, \emph{hadroquarkonium}, \emph{hybrids}, and \emph{kinematic effects} such as those arising from triangle singularities~\cite{Lebed:2016hpi}.  The paradigm to be used in this work, the \emph{dynamical diquark model}~\cite{Brodsky:2014xia, Lebed:2017min} describes exotic hadrons as diquark-antidiquark systems bound by Born-Oppenheimer (BO) potentials, and provides a natural accounting of the spectroscopy and decay patterns of multiquark exotic hadrons.

A central element of the model is its analogy with the BO treatment of heavy-quark hybrids in lattice QCD\@.  But rather than a heavy-quark pair $Q\bar Q$, the heavy diquark $\de$ and antidiquark $\bde$  pair now serve as static color sources, while the lighter, more rapidly changing gluonic field connecting them generates effective adiabatic potentials in which the $\de\bde$ pair interact.  These BO potentials, which may be computed on the lattice, then govern the dynamics of the heavy $\de\bde$ pair.  The separation of scales between the heavy diquarks and the fast light fields ($m_\de \gg \Lambda_{\rm QCD}$) provides a natural dynamical foundation for the spectroscopic study of exotics.

The model was first applied to hidden-charm exotics in Ref.~\cite{Giron:2019bcs}, successfully reproducing the observed \(\chi_{c1}(3872)\) and \(T_{c\bar c 1}\) spectrum and predicting additional states.  Refinements incorporating spin-dependent interactions and finite-size effects~\cite{Giron:2019cfc} explained fine-structure splittings and the peculiar nature of \(\chi_{c1}(3872)\).  The model was subsequently extended to $P$-wave excitations~\cite{Giron:2020fvd}, hidden-bottom and hidden-charm/strange systems~\cite{Giron:2020qpb}, all-charm tetraquarks~\cite{Giron:2020wpx}, open-strangeness candidates~\cite{Giron:2021sla}, and the open-heavy-flavor state $T_{cc}$~\cite{Lebed:2024zrp}.  Beyond tetraquarks, the same framework has been generalized (by introducing color-triplet {\em triquarks}) to pentaquarks~\cite{Lebed:2015tna} and their fine-structure spectrum~\cite{Giron:2021fnl}, consistently demonstrating that the interplay of color-triplet diquark/triquark dynamics and light gluonic fields allows the model to capture essential features of the exotic-hadron spectrum.
 
The expansion of this framework to include the close association of multiple exotic candidates with specific di-hadron thresholds (such as $\chi_{c1}(3872)$ with $D^0$-$\bar D^{*0}$, which differ in mass by only a fraction of an MeV~\cite{ParticleDataGroup:2024cfk}) mandates the generalization of the original dynamical diquark model---which may be described as \emph{adiabatic\/} because it makes use of only single BO potentials---to the \emph{diabatic\/} formalism, which provides a systematic, rigorous treatment of threshold dynamics and channel mixing.  This approach generalizes the BO approximation to describe heavy-quark systems that lie near open-heavy-flavor thresholds, accounting nonperturbatively for the mixing of such states with di-meson components.  Originally developed in atomic physics in a specific form derived in Ref.~\cite{Baer:2006}, the method was first introduced in hadronic physics to study heavy quarkonium, thus allowing a complete description of near-threshold charmoniumlike~\cite{Bruschini:2020voj,Bruschini:2021cty} and bottomoniumlike~\cite{Bruschini:2021sjh} states, as well as their scattering behavior~\cite{Bruschini:2021ckr}.  More recently, the diabatic formalism has been applied to the dynamical diquark model, where it allows for an explicit treatment of coupled-channel effects and strong mixing between $\de\bde$ and di-meson components~\cite{Lebed:2022vks,Lebed:2023kbm,Lebed:2024rsi}.

While the applications described above refer to heavy-quark systems ($c\bar c$, $b\bar b$, $cc$) for which the heavy-light scale separations are clear, it is natural to investigate the relevance of the approach for the much more marginal case of exotic hadrons containing hidden strangeness ($s\bar s$), where the scale-separation parameter $\Lambda_{\rm QCD}/m_s$ cannot be considered especially small.  In Ref.~\cite{Jafarzade:2025qvx}, we applied the dynamical diquark model to hidden-strange tetraquarks.  There, we identified $P$-wave $s\bar{s}q\bar{q}$ states with $J^{PC}=1^{--}$, $0^{-+}$, {\it etc.}, near $2.3~\text{GeV}$, $S$-wave $s\bar s s\bar s$ states with $J^{PC}=0^{++}$, $1^{+-}$, $2^{++}$ near $2.2~\text{GeV}$, and $P$-wave $s\bar s s\bar s$ states near $2.7~\text{GeV}$.  Since many of these masses lie in the vicinity of di–meson and baryon–antibaryon thresholds, it is natural to further investigate their properties within the diabatic formalism, especially to uncover whether any known states in this region appear to be heavily influenced by nearby thresholds: Are there $s\bar s$ analogues to, {\it e.g.}, $\chi_{c1}(3872)$-$D^0 \bar D^{*0}$?

In fact, experimental evidence for unconventional isoscalar resonances in the $1.8\text{--}2.6~\text{GeV}$ region already exists and continues to accumulate.  Within this mass interval, several well-established $0^{++}$ and $2^{++}$ states are reported by the Particle Data Group (PDG)~\cite{ParticleDataGroup:2024cfk}, including $f_0(1810), f_0(2020)$, $f_0(2100)$, $f_0(2200)$, $f_2(2010)$, $f_2(2150)$, $f_2(2300)$, and $f_2(2340)$~\cite{ParticleDataGroup:2024cfk}---likely more than simple $q\bar q$ models can accommodate. Among these resonances, $f_0(2020)$ and $f_2(2340)$ states exhibit the largest decay widths, $\Gamma = 440\pm 50~\text{MeV}$ and $331^{+27}_{-18}~\text{MeV}$, respectively~\cite{ParticleDataGroup:2024cfk}.
They have been prominently observed in radiative $J/\psi$ decays such as $J/\psi \to \gamma \eta'\eta'$, where partial-wave analyses by the BESIII Collaboration have established clear scalar and tensor contributions~\cite{BESIII:2022zel}.  \( f_0(2100), f_2(2010), f_2(2300)\), and \(f_2(2340)\)  were also observed by BESIII in the channel \(J/\psi\rightarrow\gamma\phi\phi\)~\cite{BESIII:2016qzq}.  In addition, the PDG lists two isoscalar $0^{-+}$ [$\eta(2225)$, confirmed by BESIII to appear in $J/\psi\rightarrow\gamma\phi\phi$~\cite{BESIII:2016qzq}, and $\eta(2370)$ (modes discussed below)] and two $1^{--}$ [$\phi(2170)$ and $\omega(2200)$] resonances. 

The resonance $\phi(2170)$ [formerly $Y(2175)$] has been reported by three independent experiments.  It was first observed by the BaBar Collaboration in the process $e^{+}e^{-}\!\to \phi f_0(980)$~\cite{BaBar:2006gsq}, and was later confirmed by the BES~\cite{BES:2007sqy}  and Belle~\cite{Belle:2008kuo} collaborations. Its mass and decay width are summarized in the PDG as $2164 \pm 5$~MeV and $88^{+26}_{-21}$~MeV, respectively.   Its nature has been discussed in a variety of theoretical frameworks.  One possibility is that it represents a hidden-strangeness tetraquark of the type $s \bar s q \bar q$ ~\cite{Agaev:2019coa}, or even a fully strange $s s \bar s \bar s$ state~\cite{Chen:2008ej}.  It has also been proposed as a $\Lambda\bar{\Lambda}$ baryonium bound state~\cite{Dong:2017rmg}.  In alternative scenarios based on hadronic molecules,   $\phi(2170)$ may be dynamically generated as a $\phi K \bar{K}$ system, where the $K \bar{K}$ pair arises from a $f_0(980)$ resonance~\cite{MartinezTorres:2008gy}.  Another recent proposal~\cite{Wei:2025ejv} identifies the peak as arising from a triangle singularity mechanism: In $e^+e^- \to \phi \pi^+ \pi^-$, the intermediate process $K_1 \bar{K}$ with $K_1 \to \phi K$, followed by $K \bar{K} \to \pi^+ \pi^-$, develops a kinematical singularity that mimics a resonance structure around $2.17~\text{GeV}$.

In the light-hadron sector, BESIII has revealed a series of unconventional resonances in radiative and hadronic $J/\psi$ and $\psi(3686)$ decays.  One of the first such states was $X(1835)$~\cite{BES:2003aic}, confirmed in $J/\psi \to \gamma \pi^+\pi^- \eta'$ with $m =1831.8^{+4.0}_{-2.6} ~\text{MeV}$ and $\Gamma =120\pm 70$~MeV~\cite{BES:2005ega}.  More recently, BESIII has expanded this spectrum: The aforementioned $\eta(2370)$ has been observed in $J/\psi \to \gamma \pi^+ \pi^- \eta^\prime$~\cite{BESIII:2010gmv}, $\gamma K^+ K^- \eta^\prime$, and $\gamma K_s^0 \bar K_s^0 \eta^\prime$~\cite{BESIII:2019wkp}, and has $J^{PC} = 0^{-+}$, $m = 2377\pm 9$~MeV, and $\Gamma = 148^{+80}_{-28} $~MeV\@.  It has been discussed as a potential glueball candidate, although lattice simulations predict the lightest $0^{-+}$ glueball state to occur near 2.6~GeV~\cite{Morningstar:1999rf,Chen:2005mg,Athenodorou:2020ani}.  A state $X(2500)$, observed in $J/\psi \to \gamma \phi \phi$ with $J^{PC}=0^{-+}$, has $m = 2470^{+15}_{-19}~\text{MeV}$ and $\Gamma = 230^{+64}_{-35}~\text{MeV}$~\cite{BESIII:2016qzq}; $X(2600)$, seen in $J/\psi \to \gamma \pi^+\pi^- \eta'$, was measured to have $m  = 2618\pm2~\text{MeV}$ and $\Gamma = 200\pm 8~\text{MeV}$ , and $J^{PC} = 0^{-+}, 2^{-+}$ are favored~\cite{BESIII:2022sfx}; and most recently, $X(2300)$ was reported in $\psi(3686) \to \phi \eta \eta'$ with $J^{PC} = 1^{+-}$, $m =  2316 \pm 31~\text{MeV}$, and $\Gamma = 89 \pm 30~\text{MeV}$~\cite{BESIII:2025prl}.  Together, these results reveal a remarkably rich spectrum of potential hidden–strangeness resonances in the region $1.8$–$2.6~\text{GeV}$ (see Fig.~\ref{fig}  at the beginning of Sec.~\ref{sec:Results}), providing an important testing ground for nonperturbative QCD dynamics.

The structure of this paper is as follows.  In Sec.~\ref{sec:States}, we enumerate the relevant tetraquark states within the dynamical diquark model.  Section~\ref{sec:Thresholds} presents the list of $s\bar s$-containing di-hadron thresholds in the range 1.8--2.6~GeV and their corresponding quantum numbers.  In Sec.~\ref{sec:Diabatic}, we review the diabatic formalism and construct the potential matrix for the hidden-strangeness sector.  Section~V presents our numerical results, analyzes the dominant di-hadron components, and compares with experimental candidates.  Finally, Sec.~\ref{sec:Concl} summarizes our conclusions and outlines directions for future study.

\section{Tetraquark States within the Dynamical Diquark Model} \label{sec:States}

In light of the quantum numbers of the candidates we have discussed in the Introduction, our focus is on tetraquark states with quantum numbers $J^{PC} = 0^{++}$, $1^{+\pm}$, $2^{++}$, $0^{-+}$, and $1^{--}$.  In this section, we briefly identify their origin within the dynamical diquark model.  We consider two types of diquarks (and their antiparticles): mixed-flavor $\delta = sq$ (where $q = u,d$) and fully strange $\delta = ss$.  Both are assumed to transform under the  antisymmetric color-$\bar{\bf 3}$ representation (being the unique attractive combination at small separations), and therefore the spin-space-flavor wave function of each diquark is completely symmetric.  The constituent quarks in each of $\de,\bde$ are assumed to be in an $S$-wave, and so the structure of each tetraquark is determined by the spin coupling within each of $\de$ and $\bde$, the total spin of the $\de\bde$ pair, and the orbital angular momentum $L$ between $\de$ and $\bde$.

For $\delta = sq$, the spectrum to be considered here includes both $S$-wave ($L=0$) and $P$-wave ($L=1$) states, although, of course, arbitrary orbital excitations are possible~\cite{Lebed:2017min}.  Labeling states by the total diquark spins $s_\de$, $s_\bde$ and total quark spin $S$ as $\left| s_\de , s_\bde \right>_S$, then in the $S$-wave sector where $J = S$, one obtains two scalar configurations with $J^{PC} = 0^{++}$, namely,
\begin{align}
X_0 = \left| 0_\de,0_\bde \right>_0 , \qquad 
X_0' = \left| 1_\de,1_\bde \right>_0 ,
\end{align}
one axial vector with $J^{PC}=1^{++}$, two with $J^{PC}=1^{+-}$,
\begin{align}
 J^{PC} = 1^{++}: & \ & X_1 = \frac{1}{\sqrt 2} \left( \left| 1_\de ,
0_\bde \right>_1 \! + \left| 0_\de , 1_\bde \right>_1 \right) \, ,
\nonumber \\
J^{PC} = 1^{+-}: & \ & Z \  = \frac{1}{\sqrt 2} \left( \left| 1_\de ,
0_\bde \right>_1 \! - \left| 0_\de , 1_\bde \right>_1 \right) \, ,
\nonumber \\
& & Z^\prime \, = \left| 1_\de , 1_\bde \right>_1 \, , \hspace{7.2em}   
\end{align}
and a tensor state $J^{PC}=2^{++}$,
\begin{align}
X_2 = |1_\delta,1_{\bar\delta}\rangle_2  \, .
\end{align}
for a total of 6 states.  If both $I \! = \! 0$ and 1 are counted, then the total becomes 12, although we are only interested in isoscalar states in this work.  Note that all such states have positive parity because they contain equal numbers of quarks and antiquarks.

In the $P$-wave $s\bar s q\bar q$ sector (again assuming the components quarks in each of $\de,\bde$ to be in an $S$-wave), the coupling of $\de,\bde$ spins to relative orbital angular momentum $L = 1$ produces only negative-parity states.  For $J^{PC} = 0^{-+}$, one obtains the states
\begin{align}
Y_0 = \frac{1}{\sqrt 2} \big( \! \left| 0_\de,1_\bde \right>_1^{L=1} \! - \left| 1_\de,0_\bde \right>_1^{L=1} \! \big) , \
Y_0' = \left |1_\de,1_\bde\right>_1^{L=1} \! ,
\end{align}
while the vector channel with $J^{PC} = 1^{--}$ contains a richer set of possibilities, namely,
\begin{align}\nonumber
Z_1 = \left|0_\de,0_\bde\right>_{0}^{L=1} \! , \quad &
Z_1' = \frac{1}{\sqrt 2} \big( \! \left| 0_\de,1_\bde \right>_1^{L=1} \! + \left| 1_\de,0_\bde \right>_1^{L=1} \! \big) , \\
Z_1'' = \left|1_\de,1_\bde \right>_0^{L=1} \! , \quad
&
Z_1''' = \left|1_\de,1_\bde\right>_2^{L=1} \! .
\end{align}
The $L=1$ $s\bar s q\bar q$ multiplet also contains states such as $J^{PC} = 0^{--}$, $2^{-+}$, {\it etc.} (for a total of 14 states~\cite{Lebed:2017min}), which we do not consider here.

The multiplets simplify considerably for $\delta = ss$.  Here, the Pauli principle forbids $s_\de = 0$, so that only $s_\de = 1$ configurations are admissible.  As a result, the number of independent states is reduced compared to the $\de = sq$ case. In the $S$-wave, the scalar and axial-vector channels are each restricted to a single state,
\begin{align} \nonumber
X_0' = \left|1_\de,1_\bde \right>_0 \, , \qquad J^{PC}=0^{++},\\
Z^\prime = \left| 1_\de , 1_\bde \right>_1 \, , \qquad J^{PC}=1^{+-},    
\end{align}
and the tensor channel remains,
\begin{align}
X_2 = \left|1_\de,1_\bde\right>_2, \qquad J^{PC}=2^{++}.
\end{align}
For the $P$-wave excitations, the pseudoscalar channel reduces  from two to one configuration,
\begin{align}
Y_0' = \left|1_\de,1_\bde\right>_1^{L=1}, \qquad J^{PC}=0^{-+},
\end{align}
while the vector channel contains only two remaining states,
\begin{align}
Z_1'' \, = \left|1_\de,1_\bde\right>_0^{L=1}, \ \  
Z_1''' = \left|1_\de,1_\bde\right>_2^{L=1}, \ \ J^{PC}=1^{--} .
\end{align}
The multiplet is completed by including states with $J^{PC} = 1^{-+}, 2^{--}$, $2^{-+}$, and $3^{--}$, for a total of 7~\cite{Giron:2020wpx}.

In summary, the restriction imposed by Fermi statistics reduces the multiplicity of states for the $\delta = ss$ sector compared to the $\delta = sq$ sector: the $0^{++}$, $0^{-+}$, and $1^{+-}$ channels each decrease from 2 to 1 independent states, and the $1^{--}$ channel decreases from 4 to 2\@.  This systematic reduction reflects the fundamental role of Fermi statistics in shaping the exotic spectrum.

\section{Hadronic Thresholds}
\label{sec:Thresholds}

In this section, we tabulate the thresholds relevant for hidden-strange tetraquarks by identifying mesons (isoscalar and open-strange) and strange baryons that have well-defined quantum numbers in pairs.  We limit the list to hadrons that contain, or potentially can contain, valence $s$ and/or $\bar s$ quarks (thus removing isovector meson pairs from consideration), and thereby in di-hadron pairs can contribute to $s\bar s q\bar q$ or $s\bar s s\bar s$ states.  We furthermore limit the list to include only those hadrons with $\Gamma \lesssim 100$~MeV, since identifying a di-hadron component within a tetraquark requires each constituent to be at least as stable as the overall state to which it contributes.  A tetraquark with two $\Gamma = 100$~MeV constituents that can decay independently would thus have $\Gamma \gtrsim 200$~MeV and might be difficult to discern in the data.  Additionally, the precise threshold for such broad constituent states becomes a rather ill-defined concept.

Each hadron is specified by its total spin \(J\), parity \(P\), charge-conjugation parity \(C\) (if applicable), and internal orbital angular momentum \(L\).  For neutral \(q\bar{q}\) mesons, these quantum numbers satisfy the relations:
\begin{align}
P = (-1)^{L+1}, \qquad C = (-1)^{L+S}\,.
\end{align}
Here, \(S\) is the total spin of the $q\bar q$ pair, and \(L\) is their relative orbital angular momentum,  which also give rise to the usual spectroscopic notation \(n\,^{2S+1}L_J\).

The list of these mesons, including $J^{PC}$ quantum numbers, masses, decay widths, and \(n\,^{2S+1}L_J\) values inferred by the PDG~\cite{ParticleDataGroup:2024cfk}, is presented in Table~\ref{tab:mesonspec}\@.  As noted above, we restrict component hadrons to relatively narrow states in order to ensure that they produce well-defined thresholds in the coupled-channel framework.  In particular, broad isoscalar resonances such as \(f_2(1270)\) and \(h_1(1170)\) are excluded from our list because their decay widths exceed several hundred MeV, making them unsuitable for our purpose.  In Table~\ref{tab:mesonspec} we also include the strange mesons \(K^*(892)\) and \(K_1(1270)\), together with the baryons \(\Lambda(1116)\) and \(\Xi(1315)\).  Unlike the isoscalar mesons, these hadrons are not individually $C$ eigenstates, but when combined into particle-antiparticle pairs (corresponding to the threshold states), they may be expressed as states of definite $C$ eigenvalue.

\begin{table}[h]
\caption{Masses, widths (central values, in MeV), and quantum numbers $J^{PC}$ of the hadrons used in this work, and spectroscopic assignments $n\,{}^{2S+1}L_J$ as inferred by the PDG~\cite{ParticleDataGroup:2024cfk}.  For states with nonzero isospin,  measured values refer to the lighter isobar.}
\label{tab:mesonspec}
\setlength{\extrarowheight}{1.5ex}
\begin{tabular*}{\columnwidth}{@{\extracolsep{\fill}}lcccc@{}}
\hline\hline
Hadron & Mass & Width & $J^{PC}$ &  $n\,{}^{2s+1}L_J$ \\
\hline
$\eta(548)$       & 547.9   & $1.31 \times 10^{-3}$     & $0^{-+}$   & $1\,{}^1S_0$ \\
$\eta (958)$& 957.8   & 0.188    & $0^{-+}$   & $1\,{}^1S_0$ \\
$\eta(1295)$      & 1294    & 55       & $0^{-+}$   & $2\,{}^1S_0$ \\
$\eta(1475)$      & 1476    & 96       & $0^{-+}$   & $2\,{}^1S_0$ \\
$\omega(782)$     & 782.7   & 8.49     & $1^{--}$   & $1\,{}^3S_1$ \\
$\phi(1020)$      & 1019.5  & 4.25     & $1^{--}$   & $1\,{}^3S_1$  \\

$h_1(1415)$       & 1409    & 78       & $1^{+-}$   & $1\,{}^1P_1$ \\
$f_1(1285)$       & 1281.8  & 23.0     & $1^{++}$   & $1\,{}^3P_1$ \\
$f_1(1420)$       & 1428.4  & 56.7     & $1^{++}$   & $1\,{}^3P_1$ \\
$f_2(1525)$       & 1517.3  & 72       & $2^{++}$   & $1\,{}^3P_2$  \\
$K^*(892)$        & 890     & 51.4     & $1^{-}$    & $1\,{}^3S_1$ \\
$K_1(1270)$       & 1253    & 101      & $1^{+}$    & $^1P_1 \,$-$^3P_1$ \\
$\Lambda$         & 1115.7    & $\sim 0$       & $\tfrac{1}{2}^+$ & $1\,{}^1S_0$ \\
$\Xi$             & 1314.8    & $\sim 0$       & $\tfrac{1}{2}^+_{\vphantom{y}}$ & $1\,{}^1S_0$ \\
\hline\hline
\end{tabular*}
\end{table}

To classify the quantum numbers of the di-hadron thresholds, we follow the formalism of Ref.~\cite{Bruschini:2020voj}.  Consider two hadrons with intrinsic spins \(S_1\) and \(S_2\).  Their total spin is obtained by coupling 
\(
\mathbf{S} = \mathbf{S}_1 + \mathbf{S}_2,
\)
while the eigenvalue of the relative orbital angular momentum {\bf L} between the two hadrons (the partial wave) is \(\ell\).  The total angular momentum of the two-hadron system is then
\(
\mathbf{J} = \mathbf{L} + \mathbf{S}.
\)
The parity of the two-hadron state is
\(
P = P_1 P_2 (-1)^{\ell},
\)
where \(P_1, P_2\) are the intrinsic parities of the individual hadrons.  A baryon-antibaryon pair has $P_1 P_2 = -1$.  For $C$-parity, one must distinguish two cases:
\begin{itemize}
    \item For either identical mesons (\(M_1 = M_2\)) or baryons (\(B_1 = B_2\)) [generically, hadrons $H_1 = H_2$], the $C$-parity of the $H_1 {\bar H}_2$ di–hadron component is well-defined:
\begin{align}
C = (-1)^{\ell+S}.
\end{align}
\item For nonidentical isoscalar mesons (\(M_1 \neq M_2\)), or for open-strange mesons such as kaons, or for non-identical baryons (not used in this work), states with either $C = +1$ or $-1$ can be constructed.
\end{itemize}
  
All relevant quantum numbers for the meson (baryon) thresholds are listed in Tables~\ref{tab:meson_thresholds-qs}--\ref{tab:ss-Meson thresholds}.

\begin{table}[h!]
\caption{Di-meson pairs $M_1 \bar{M}_2$ (threshold energy $T$ in GeV) and relative orbital angular momentum values $\ell$ that allow 
$S$-wave $s\bar{s}q\bar{q}$ candidates for the given $J^{PC}$ quantum numbers.}
\label{tab:meson_thresholds-qs}
\centering    
\setlength{\extrarowheight}{1.5ex}
\begin{tabular*}{\columnwidth}{@{\extracolsep{\fill}}lcccc@{}}
\hline\hline
\textbf{$M_1 \bar{M}_2$} & $T_{M_1\bar{M}_2}$ & 
\multicolumn{3}{c}{$\ell$ for $J^{PC}$} \\
\cline{3-5}
 &  & $0^{++}$ &  $1^{++}$&  $2^{++}$  \\
\hline
\(\eta \, \eta(1295) \)       & 1.842  &  0 &   -- &  2    \\
\(\eta(958)\eta(958)\)           & 1.916 &  0 &  -- &  2   \\
\(\eta h_1(1415)\)     & 1.957 & 1 &  1 & 1,3   \\
\(\eta f_1(1420)\)     & 1.976 &  1 &  1 &  1,3   \\
\(\eta(958) \phi\)     & 1.977 &  --  &  0,2 &  2     \\
\(\eta \, \eta(1475)\)     & 2.024 &  0 &  -- &  2    \\
\(\phi \phi\) & 2.039  &  0,2 &   2 &  0,2,4    \\
\(\eta f_2(1525)\)     & 2.065 &  -- &  1,3 &  1,3    \\
 
\(\omega \eta(1295)\) & 2.077  &   -- &  0,2  &   2     \\ 
\(K^* \bar{K}_1(1270)\) & 2.143 &  1 &  1,3 &  1,3    \\
\hline\hline
\end{tabular*}
\end{table}

\begin{table}[h]
\caption{Di-hadron pairs $H_1 \bar{H}_2$ (threshold energy $T$ in GeV) and relative orbital angular momentum values \(\ell\) that allow $P$-wave $s\bar{s}q\bar{q}$ and $S$-wave $s\bar{s}s\bar{s}$ candidates for the given \(J^{PC}\) quantum numbers.}
\label{tab:meson_thresholds-mixed}
\centering
\setlength{\extrarowheight}{1.5ex}
\begin{tabular*}{\columnwidth}{@{\extracolsep{\fill}}lcccccc@{}}
\hline\hline
\textbf{$H_1 \bar{H}_2$} & $T_{H_1\bar{H}_2}$ & 
\multicolumn{4}{c}{$\ell$ for $J^{PC}$} \\
\cline{3-7}
 &  & $0^{++}$ & $1^{+-}$  & $2^{++}$ & $0^{-+}$ & $1^{--}$ \\
\hline
\(\Lambda\bar{\Lambda} \)       & 2.231 &  1 &  1 &  1,3  & 0 & 0,2 \\
\(\eta(958) f_1(1285)\) & 2.240 &  1 &  1 &  1,3 & -- & 0,2 \\
\(\eta(958) \eta(1295)\) & 2.252 &   0 &  -- &  2 & -- & 1  \\
\(\omega \eta(1475)\) & 2.259  & -- &  0,2  &   2 & 1 & 1  \\
\(\omega f_2(1525)\) & 2.300 &  1,3  & 1,3  & 1,3,5 & 2  & 0,2,4 \\
\(\phi f_1(1285)\)     & 2.301&  1  &  1,3  &  1,3  & 0,2 & 0,2\\
\(\eta(1295)\phi \) & 2.313 &   -- &  0,2  &   2  &1 & 1 \\
 $\eta(958)\,h_1(1415)$            & 2.367 &  1 &  1 &  1,3 & -- & 0,2\\
\(\eta(958)f_1(1420)\)           & 2.386 & 1 &  1 &  1,3 & -- &0,2 \\
\(\phi h_1(1415)\)           & 2.428 & 1 &  1,3 &  1,3  & 0,2 & 0,2 \\
\(\eta(958)\eta(1475)\)           & 2.434 &  0 &  -- &  2  & -- & 1  \\ 
\(\phi f_1(1420)\)           & 2.448 & 1  &  1,3  &  1,3  & 0,2  & 0,2 \\
\(\eta(958) f_2(1525)\) & 2.475 &   -- &  1,3 &  1,3  & 2 & 2  \\
\( \eta(1475)\phi\)           & 2.495 & -- &  0,2  &   2  & 1 & 1 \\
\(K_1(1270) \bar{K}_1(1270)\) & 2.506 & 0,2 & 2 & 0,2,4 & 1& 1,3 \\
\(f_1(1285)f_1(1285)\) & 2.564& 0,2 & 2 & 0,2,4 &1 & 1,3  \\
\(f_1(1285)\eta(1295)\) & 2.576& 1 & 1 & 1,3 & -- &  0,2  \\
\(\eta(1295)\eta(1295)\) & 2.588&  0 & -- & 2 & -- & 1  \\
\hline\hline
\end{tabular*}
\end{table}

\begin{table}[h]
\caption{Di-hadron pairs $H_1 \bar{H}_2$ (threshold energy $T$ in GeV) and relative orbital angular momentum values \(\ell\) that allow $P$-wave $s\bar{s}s\bar{s}$ candidates for the given \(J^{PC}\) quantum numbers.}
    \label{tab:ss-Meson thresholds}
\centering
\setlength{\extrarowheight}{1.5ex}
    \begin{tabular*}{\columnwidth}{@{\extracolsep{\fill}}cccc@{}}
\hline\hline
\textbf{$H_1 \bar{H}_2$} & $T_{H_1\bar{H}_2}$ [GeV] & 
\multicolumn{2}{c}{$\ell$ for $J^{PC}$} \\
\cline{3-4}
 &    & $0^{-+}$ & $1^{--}$ \\
\hline
         \(\Xi\bar{\Xi} \)       & 2.630   & 0 &  0,2  \\
\(f_1(1285) h_1(1415)\) & 2.691    & 1 &  1,3   \\
\(\eta(1295) h_1(1415)\) & 2.703     & -- &  0,2   \\
\(f_1(1285) f_1(1420)\) & 2.710     & 1 &  1,3  \\
\(\eta(1295) f_1(1420)\) & 2.722     & -- &  0,2   \\
\(\eta(1295) f_2(1525)\) & 2.811     & 2 &  2   \\
\(h_1(1415)h_1(1415)\) & 2.818 & 1 & 1,3   \\
\(h_1(1415) f_1(1420)\)     & 2.837  & 1 &  1,3   \\
\(f_1(1420)f_1(1420)\) & 2.857  & 1 & 1,3  \\
\(\eta(1475)h_1(1415)\)     & 2.885   & -- &  0,2  \\
\(\eta(1475)f_1(1420)\)     & 2.904   & -- &  0,2    \\
 \(f_2(1525)h_1(1415)\)     & 2.926    & 1,3 &  1,3  \\
\hline\hline
\end{tabular*}
\end{table}

\section{Diabatic Description of the Dynamical Diquark Model}
\label{sec:Diabatic}

The original dynamical diquark model for exotic hadrons describes tetraquarks as bound states of a diquark–antidiquark pair $\de\bde$ interacting through gluonic field configurations~\cite{Giron:2019bcs} (and pentaquarks as analogous diquark-triquark bound states~\cite{Lebed:2015tna}).  The natural framework for this configuration is the Born-Oppenheimer (BO) approximation: The heavy $\de$ and $\bde$ move slowly compared to the light fields, which adjust instantaneously (``adiabatically'') to the motion of $\de$ and $\bde$.  The $\de\bde$ system with reduced mass $\mu_{\de\bde}$ then satisfies
\begin{equation} \label{eq:SE}
\left[ \, \frac{\mathbf{p}^2}{2 \mu_{\de\bde} }+ H_{\rm light}(\mathbf{r}) - E \, \right] |\psi
\rangle = 0\,,
\end{equation}
where $H_{\rm light}(\mathbf{r})$ encodes the Hamiltonian of the light-field degrees of freedom (d.o.f.) at a particular diquark
separation $\mathbf{r}$.  Solving
\begin{align}
H_{\rm light}(\mathbf{r})|\xi_i(\mathbf{r})\rangle= V_i(\mathbf{r})\,|\xi_i(\mathbf{r}) \rangle\,,
\end{align}
defines the adiabatic potentials $V_i(\mathbf{r})$, which have been calculated in lattice-QCD simulations~\cite{Juge:1997nc,Juge:1999ie,Juge:2002br} as the potential-energy functions obtained between any two static color-triplet sources at separation $\mathbf{r}$.  Expanding the full wave function within this basis gives
\begin{equation} 
|\psi \rangle = \sum_{i} \int d\mathbf{r}\, \tilde \psi_i
(\mathbf{r} ) \, |\mathbf{r}  \rangle \,
|\xi_i(\mathbf{r} ) \rangle,
\end{equation}
where $\left| \mathbf{r} \right>$ denotes the state corresponding to the two heavy sources with separation vector $\mathbf{r}$.  This expression is called the {\it adiabatic expansion}, because it suggests that the light d.o.f.\ remain in particular states $\left| \xi_i (\textbf{r}) \right>$ that respond only to the heavy-source separation $\mathbf{r}$.  In general, however, different light-d.o.f.\ states can develop coupled-channel mixing when $\mathbf{r}$ changes:
\begin{align}
 \langle \xi_j(\mathbf{r})|\nabla \xi_i(\mathbf{r})\rangle \neq 0 \, .
\end{align}
These overlaps are called {\it nonadiabatic couplings}.  If they can be neglected, then one obtains the single-channel BO-approximation Schr\"{o}dinger equation:
\begin{equation} 
  \left[ - \frac{\hbar^2}{2 \mu_{\delta\bar{\delta}}}    \nabla ^2 +
V_{i}(\mathbf{r})-E   \right] \tilde \psi_i (\mathbf{r}) = 0 \, .
\end{equation}

Near di-meson thresholds, however, mixing between $\de\bde$ and di-meson $M_1 \overline{M}_2$ configurations becomes essential, invalidating this approximation.  A more general treatment employs the \emph{diabatic\/} expansion, in which the basis states are defined at a fiducial separation $\mathbf{r}_0$, whose value  ensures that $V_i(\mathbf{r}_0)$ lies far from thresholds.  $\mathbf{r}_0$ is chosen so that the states $\left| \xi_i (\mathbf{r}_0) \right>$ can reliably be expressed in terms of unmixed configurations.  The wave function is then
expanded as
\begin{equation} 
|\psi \rangle = \sum_{i} \int d\mathbf{r}' \, \tilde \psi_i(\mathbf{r}',\mathbf{r}_0) \,
|\mathbf{r}' \rangle \, |\xi_i(\mathbf{r}_0) \rangle \, ,
\end{equation}
leading to the coupled-channel Schrödinger equation
\begin{equation} 
\sum_{i} \left[ - \frac{\hbar^2}{2 \mu_{i}} \de_{ij}  \nabla ^2 +
V_{ji}(\mathbf{r}, \mathbf{r}_0)-E \de_{ji} \right] \psi_i (\mathbf{r}, \mathbf{r}_0) = 0 \, ,
\end{equation}
where the {\it diabatic potential matrix\/} is defined by
\begin{equation}
V_{ji}(\mathbf{r}, \mathbf{r}_0) \equiv \langle \xi_j(\mathbf{r}_0) | H_{\rm light}(\mathbf{r}) | \xi_i(\mathbf{r}_0) \rangle \, .
\end{equation}

With the argument $\mathbf{r}_0$ suppressed from this point on, the diagonal elements $V_{ii}(\mathbf{r})$ describe uncoupled-channel potentials,
while the off-diagonal terms $V_{ji}(\mathbf{r})$ represent physical mixing between
configurations.  Threshold effects are naturally included: Closed-channel (below-threshold) wave functions decay
exponentially, while open-channel (above-threshold) wave functions support oscillatory asymptotics and resonance
behaviors~\cite{Bruschini:2021cty}.  The diabatic formalism  thereby provides a
systematic description of strongly coupled multichannel systems, and thus is well suited to studies of exotic hadrons~\cite{Lebed:2023kbm}.

The coupled equations can be written compactly as
\begin{equation}
 \big[ \, \mathbf{K}+\mathbf{V}(\mathbf{r})-E \, \big] \, \mathbf{\Psi}(\mathbf{r})=0,
\end{equation}
where the kinetic-energy operator is
\begin{equation}
     \mathbf{K}=
     \begin{pmatrix}
         -\dfrac{\hbar^{2}}{2\mu_{\delta\bar \delta}}\nabla^{2}&\\
         &-\dfrac{\hbar^2}{2\mu_{M_1\overline{M}_2}^{(1)}}\nabla^2\\
         & &\ddots\\
         & & &-\dfrac{\hbar^2}{2\mu_{M_1\overline{M}_2}^{(N)}}\nabla^2
     \end{pmatrix},
\end{equation}
and the coupled wave function is
\begin{equation}
     \mathbf{\Psi}(r)=
     \begin{pmatrix}
         \psi_{\delta\bar{\delta}}(r)\\
         \psi_{M_1\overline{M}_2}^{(1)}(r)\\
         \vdots\\
         \psi_{M_1\overline{M}_2}^{(N)}(r)
     \end{pmatrix}.
\end{equation}

The specific phenomenological form that we assume for the diabatic potential matrix, which has been used since its first hadronic applications~\cite{Bruschini:2020voj,Lebed:2022vks}, is
\begin{equation}
\mathbf{V}(\mathbf{r})=
\begin{pmatrix}
V_{\de \bde}(r) & V_{\rm mix}^{(1)}(r)  & \cdots &
V_{\rm mix}^{(N)}(r) \\
V_{\rm mix}^{(1)}(r) & 
V_{M_1 \bbar M_2}^{(1)}(r) & & \\
\vdots & & \ddots & \\
V_{\rm mix}^{(N)}(r) & & & V_{M_1 \bbar M_2}^{(N)}(r) \\
\end{pmatrix},
\end{equation}
where blank elements are taken to be zero, and
\begin{align}
V_{\delta\bar{\delta}}(r) &= -\frac{\alpha}{r} + \sigma r + V_0 , \\
V_{ii}(r) &=V_{M_1 \bbar M_2}^{(i)}(r)= M_{1}^{i} + M_{2}^{i} \equiv T^{(i)}_{M_1\bar{M}_2}, \\
V_{0i}(r) &= \frac{\Delta}{2}
\exp \! \left\{ -\frac{1}{2} \frac{\left[
V_{\de \bde}(r) -
T_{M_1 \bbar M_2 }^{(i)} \right]^2}{(\sigma \rho)^2} \right\},
\end{align}
where $\alpha=0.053~\text{GeV}\cdot\text{fm}$, $\sigma=1.097~\text{GeV/fm}$,
$V_0=-0.380~\text{GeV}$ is taken from \cite{Juge:1997nc}, $\rho = 0.3~\text{fm}$, and $\Delta = 0.130~\text{GeV}$~\cite{Bruschini:2020voj}.  Note also the disappearance of any direction dependence from this application ($\mathbf{r} \to r$).

The use of a uniform mixing potential $V_{0i}(r)$ for all di-meson states $i$ (as well as the specific Gaussian functional form) is of course a substantial phenomenological assumption.  In the heavy-quark case at least, one can make headway by comparing the differing effects of thresholds in which the component hadrons are related by heavy-quark symmetry, such as $D$ and $D^*$~\cite{Bruschini:2023zkb,Lebed:2025eqm}.

\section{Results}
\label{sec:Results}

\begin{figure*}[]
    \centering
    \includegraphics[scale=0.7]{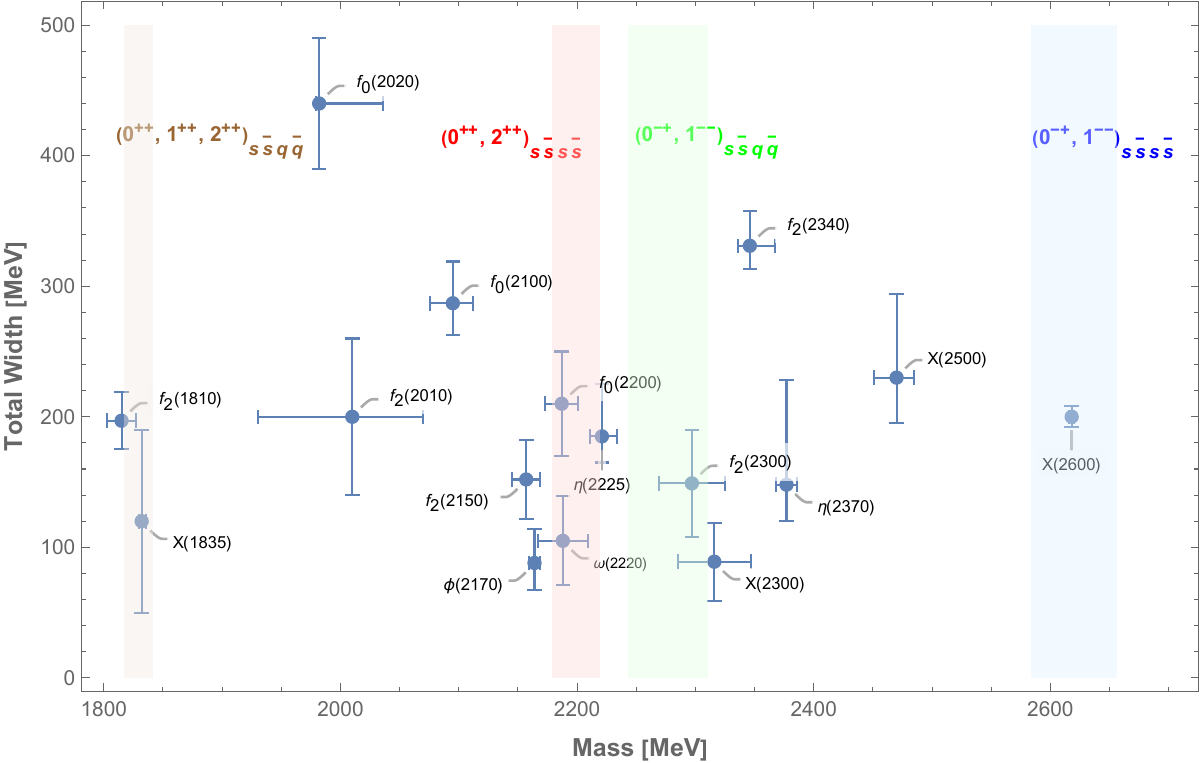}
    \caption{Spectrum of $s\bar s s\bar s$ and $s\bar s q\bar q$ states of given quantum numbers within the dynamical diquark model, expressed such that the differing di-hadron threshold effects for each state (but not yet including fine-structure effects in the underlying states) produce a spread for each multiplet indicated by a shaded band.  These results are compared to the measured values of the isoscalar, light-unflavored meson resonances (masses and decay widths) tabulated by the PDG and by BESIII.}
    \label{fig}
\end{figure*}

The inclusion of hadronic thresholds in the diabatic dynamical diquark model for hidden-strangeness tetraquarks typically leads to a reduction of the mass eigenvalues by a few tens of MeV compared to those in the adiabatic case~\cite{Jafarzade:2025qvx}.  In general, these effects become negligible once the difference between the adiabatic tetraquark mass eigenvalue and the nearest meson threshold exceeds about \(200\)~MeV\@.  In the following subsections we present results for $S$- and $P$-wave states, and for $s\bar s q\bar q$ and $s\bar s s\bar s$ quark contents, and compare these results to the spectrum of observed resonances discussed in the Introduction and displayed in Fig~\ref{fig}. 

 Across all states listed in the  following subsections, the compact  $\de\bde$ configuration is seen to be the main component; despite some mixing with nearby di‑meson thresholds, no state is threshold‑dominated. Threshold effects are slightly more pronounced for $P$-wave than for $S$-wave states, producing somewhat larger di-hadron admixtures and mass shifts, but the  $\de\bde$ component continues to dominate the overall wave function.

\subsection{$S$-Wave \(s\bar{s}q\bar{q}\) States}

For the \textit{$S$-wave \(s\bar{s}q\bar{q}\)} system, we consider thresholds starting at \(1840\)~MeV, as motivated by our previous predictions from the dynamical diquark model~\cite{Jafarzade:2025qvx}.  As seen in Table~\ref{tab:$S$-wave-qs}, the \(\delta\bar{\delta}\) configuration dominates the \(1^{++}\) and \(2^{++}\) states, each with contributions above \(90\%\).  For the \(0^{++}\) state, the \(\delta\bar{\delta}\) structure remains the primary component (\( \approx 87\%\)), while the \(\eta\eta(1295)\) channel provides the second-largest contribution (\( \approx 8\%\)).  Note that, according to Table~\ref{tab:meson_thresholds-qs}, thresholds with \( J^{PC} = 2^{++}\) occur for every listed di–meson threshold; however, the contribution from each of them is found to be $< 1$\%.

\begin{table}[h]
\caption{Diabatic formalism results for $S$-wave $s\bar s q\bar q$ states:  energy eigenvalues $E$, state sizes $\left< r \right>$, and diquark–antidiquark and di–meson state contents (\%).  Entries $< 1\%$ are left blank; “–” indicates forbidden quantum numbers.}
\label{tab:$S$-wave-qs}
\renewcommand{\arraystretch}{1.2}  
\begin{tabular*}{\columnwidth}{@{\extracolsep{\fill}}lrrr@{}}
\hline\hline
$J^{PC^{\vphantom{\dagger}}}$  & $0^{++}$ & $1^{++}$ & $2^{++}$ \\
\hline 
$E$ (GeV) & 1.817 & 1.826 & 1.822  \\
 
$\langle r \rangle$ (fm) & 0.635 & 0.570 & 0.572  \\
 
$\delta\bar{\delta}$ & 86.5  & 96.0  & 96.2   \\
 
$\eta\eta(1295)$        & 8.4  & –   &       \\
$\eta(958)\eta(958)$    & 2.4   & –   &      \\
\(\eta(958)\phi\)       & –   & 1.4  &        \\
\hline\hline
\end{tabular*}
\end{table}

The calculated sizes $\left< r \right>$ of these states are all significantly smaller that \( 1\)~fm,  suggesting a compact tet\-ra\-quark structure rather than that of a loosely bound di-meson molecule.  These results reinforce the interpretation that the states are dominated by a genuine \(\de\bde\) configuration.

The only observed light, unflavored meson states listed by the PDG found to be compatible with the mass eigenvalues of Table~\ref{tab:meson_thresholds-qs} are \(f_2(1810)\) ($2^{++}$) and \(X(1835)\) (\(0^{-+}\)).  Due to its $P = -$ eigenvalue, $X(1835)$ is incompatible with an $S$-wave tetraquark configuration.   $f_2(1810)$, however, serves as a potential $1S$ $s\bar s q\bar q$ candidate (Indeed, it possesses $K\bar K$ and $\eta \eta$ decay modes); and if so, then $0^{++}$ and $1^{++}$ partners should be observable nearby. 

Results for the $0^{++}$ and $2^{++}$ states (with $m_{\delta}=0.710$ MeV) lie well below the experimental masses of $f_2(2010)$ and $f_0(2020)$ (see Fig.~\ref{fig}). However, both of these states have been interpreted as tetraquark candidates in various theoretical works.  In particular, QCD sum-rule analyses of fully strange configurations ($s\bar{s}s\bar{s}$) assign $f_2(2010)$ to a $J^{PC}=2^{++}$ tetraquark state~\cite{Dong:2023,Su:2022eun}, while constituent-quark-model studies of light-tetraquark spectroscopy suggest that $f_0(2020)$ may correspond to a radial excitation within a scalar tetraquark nonet~\cite{Zhao:2021jss}.

\subsection{$P$-Wave \(s\bar{s}q\bar{q}\) States}

For the \textit{$P$-wave \(s\bar{s}q\bar{q}\)} system, our previous adiabatic results~\cite{Jafarzade:2025qvx} suggest that the relevant thresholds start at \(2.3\)~GeV, corresponding to the \(\omega f_{2}(1525)\) channel (see Table~\ref{tab:meson_thresholds-mixed}).  Table~\ref{tab:$P$-wave-qs} shows that the \(\delta\bar{\delta}\) configuration again dominates, contributing \( \approx 75\%\) for the \(1^{--}\) state and \( \approx 81\%\) for the \(0^{-+}\) state.

\begin{table}[h]
\caption{Diabatic formalism results for $P$-wave $s\bar s q\bar q$ states: energy eigenvalues $E$, state sizes $\left< r \right>$, and diquark–antidiquark and di–meson state contents (\%).  Entries $< 1\%$ are left blank; “–” indicates forbidden quantum numbers.
}
\label{tab:$P$-wave-qs}
    \renewcommand{\arraystretch}{1.2}
    \begin{tabular}{l@{\hskip 4em}r@{\hskip 4em}r}
%{\columnwidth}{@{\extracolsep{\fill}}lcc@{}}
\hline\hline
$J^{PC^{\vphantom{\dagger}}}$ & $1^{--}$  & $0^{-+}$\\
\hline
$E$ (GeV) & 2.243 & 2.265\\
$\langle r \rangle$ (fm)  & 0.965   & 0.936\\
$\delta\bar{\delta}$ & 75.4  &80.7 \\

\(\omega f_2(1525)\)  & 6.9  &     \\
\(\phi f_1(1285)\)   & 7.0   & 11.7 \\
 \(\eta(1295)\phi \)   &2.5   & 2.7     \\
$\eta(958)\,h_1(1415)$            & 2.0  & -   \\
\(\eta(958)f_1(1420)\)  & 1.3    &-   \\
\(\phi h_1(1415)\) & 1.1   & 1.5   \\
\(\eta(958)\eta(1475)\)           &  & -\\
\(\phi f_1(1420)\)           &    &1.2 \\
\hline\hline
\end{tabular}
\end{table}

In the \(1^{--}\) channel, the largest subleading contributions come from the \(\omega f_{2}(1525)\) and \(\phi f_{1}(1285)\) thresholds, each \( \approx 7\%\).  Referring to Table~\ref{tab:meson_thresholds-mixed}, this result is natural since both channels lie close in mass and contribute to the same lowest partial waves \((\ell =0,2)\).  For the \(0^{-+}\) channel, the largest subleading component is \(\phi f_{1}(1285)\) (\( \approx 12\%\)).  The \(\omega f_{2}(1525)\) channel contributes only weakly, as it appears only in the \(\ell=2\) partial wave.

The states \(\omega(2220)\) ($1^{--}$) and \(\eta(2225)\) ($0^{-+}$) listed in the PDG have $J^{PC}$ quantum numbers and mass eigenvalues compatible with those in Table~\ref{tab:$P$-wave-qs}\@.

\subsection{$\bm{\phi(2170)}$ and $\bm{\eta(2370)}$} \label{sec:phi}

 The different interpretations of $\phi(2170)$ can be understood as complementary perspectives applied at different levels of physical description.  In the diabatic diquark model employed here, $\phi(2170)$ is a predominantly compact $s\bar s q \bar q$ state composed of diquarks, decaying into meson pairs such as $\phi \eta$, $\omega \eta(958)$, and $K \bar K^*$ via the fall-apart mechanism.   To expand upon the discussion in the Introduction, in hadronic-molecule models such as in Ref.~\cite{MartinezTorres:2008gy}, $\phi(2170)$ emerges dynamically from meson–meson interactions ({\it e.g.}, $\phi + K\bar K$), leading to comparable final states, although the system is more spatially extended.  In the $\Lambda\bar \Lambda$ baryonium scenario \cite{Dong:2017rmg}, $\phi(2170)$ is a loosely bound baryon–antibaryon system, favoring open-strange decays such as $K \bar K$, and differing in both structure and dominant decay channels from the tetraquark picture. Finally, the triangle-singularity mechanism in Ref.~\cite{Wei:2025ejv} provides a purely kinematic explanation of the observed peak, producing similar invariant-mass enhancements without requiring a  conventional bound state.

 In the dynamical diquark model, the inclusion of spin- and isospin-dependent interactions for the $P$-wave tetraquarks~\cite{Giron:2020fvd} yields the following Hamiltonian:
\begin{eqnarray}
H & = & H_0 + 2 \kappa_{qs} ({\bf s}_q \! \cdot \! {\bf s}_s +
{\bf s}_{\bar q} \! \cdot \! {\bf s}_{\bar s}) + V_{LS} \,
{\bf L} \cdot {\bf S} \nonumber \\ & & + V_I \, {\bm \tau}_q
\! \cdot \! {\bm \tau}_{\bar q} \; {\bm \sigma}_q \! \cdot \!
{\bm \sigma}_{\bar q} + V_T \, {\bm \tau}_q \! \cdot
\! {\bm \tau}_{\bar q} \; S_{12}^{(q\bar{q})} \, .
\label{eq:FullHam}
\end{eqnarray}
Here, $H_0$ denotes the multiplet-average Hamiltonian  (with eigenvalue $M_0$), prior to the introduction of spin- and isospin-dependent fine structure. The $\de$ and $\bde$ spin operators are defined as ${\bf s}_\delta \equiv {\bf s}_q + {\bf s}_s$ and ${\bf s}_{\bar\delta} \equiv {\bf s}_{\bar q} + {\bf s}_{\bar s}$, respectively. The total quark-spin operator is given by ${\bf S} \equiv {\bf s}_\delta + {\bf s}_{\bar\delta}$, and ${\bf L}$ represents the orbital angular momentum operator between $\de$ and $\bde$.  The isospin-dependent interaction (analogous to pion exchange), here labeling  its coefficient as $V_I$, is represented in the fourth term, and the last term contains the tensor operator $S_{12} $, defined as:
\begin{equation}
\label{eq:Tensor}
S_{12} \equiv 4 \left[ 3 \, {\bf S}_1 \! \cdot {\bm r} \, {\bf S}_2 \! \cdot {\bf r} / r^2 - {\bf S}_1 \! \cdot {\bf S}_2 \right] \, ,
\end{equation}
where spin operators ${\bf S}_i$ are used so that this operator can be applied to any component particles $1,2$, including individual quarks or diquarks.  In the case of \(1^{--}\) and \(0^{-+}\) states, the explicit mass expressions~\cite{Giron:2020fvd} read:
\begin{widetext}
\begin{eqnarray}
M_{1^{--}}^{I=0}&=& M_0 \begin{pmatrix} 1 & 0 & 0 & 0\\ 0 & 1 & 0 &
0\\ 0 & 0 & 1 & 0\\ 0 & 0 & 0 & 1
\end{pmatrix}
+\kappa_{qs}
\begin{pmatrix}
0 & -\sqrt{3} & 0 & 0 \\
-\sqrt{3} & -2 & 0 & 0 \\
0 & 0 & -1 & 0 \\
0& 0 & 0 & 1 \\
\end{pmatrix}
-V_{LS}
\begin{pmatrix}
0 & 0 & 0 & 0\\
0 & 0 & 0 & 0\\
0 & 0 & 1 & 0\\
0 & 0 & 0 & 3
\end{pmatrix}
-3V_I
\begin{pmatrix}
-3 &0 & 0 & 0 \\
0 & 1 & 0 & 0 \\
0 & 0 & 1 & 0 \\
0 & 0 & 0 & 1 \\
\end{pmatrix}\nonumber\\
&&-3V_T
\begin{pmatrix}
0 & 0 & 0 & 0\\
0 & 0 & 0 & -\frac{4}{\sqrt{5}} \\
0 & 0 & -1 & 3\sqrt{\frac{3}{5}} \\
0 &-\frac{4}{\sqrt{5}}& 3\sqrt{\frac{3}{5}} & -\frac{7}{5} \\
\end{pmatrix} \, , \label{eq:1--} \\
M_{0^{-+}}^{I=0} &=& M_0\begin{pmatrix}
1 & 0 \\
0 & 1 
\end{pmatrix}
+\kappa_{qs}
\begin{pmatrix}
0 & 1 \\
1 & 0 
\end{pmatrix}
-2V_{LS}
\begin{pmatrix}
1 & 0 \\
0 & 1 \\
\end{pmatrix}-3V_{I}
\begin{pmatrix}
-3 & 0 \\
0 & 1 \\
\end{pmatrix}
+12V_T 
\begin{pmatrix}
1 & 0 \\
0 & 0 \\
\end{pmatrix}\,. \label{eq:0-+}
\end{eqnarray}
\end{widetext}

By using the results in Table IV of Ref.~\cite{Jafarzade:2025qvx} for the mass eigenvalues of the $1^{--}$ and $0^{-+}$ $s\bar s s\bar s$ states in terms of $\kappa_{ss}$, $V_{LS}$, and $V_T^\prime$ [the latter being defined as the coefficient of the tensor operator $S^{\de\bde}_{12}$ in the $s\bar s s\bar s$ Hamiltonian that is analogous to $S^{q\bar q}_{12}$ in Eq.~(\ref{eq:FullHam})], and fitting to results obtained from QCD sum rules in Ref.~\cite{Su:2022eun}, we determine the spin-dependent parameters to be $\kappa_{ss} = 16 \pm 7$~MeV, $V_{LS} = 28 \pm 5$~MeV , and $V_T^\prime = -1.5 \pm 1.6~\mathrm{MeV}$\@. Note that this value of $\kappa_{ss}$ is also consistent with ones used in Refs.~\cite{Karliner:2014gca, Maiani:2022psl}, and $V'_T$ is consistent with zero. Assuming that these parameters remain the same in the $s\bar s q \bar q$ sector ({\it i.e.}, $\kappa_{sq} \approx \kappa_{ss}$, the same value of $V_{LS}$ applies, and $V_T$, which corresponds to a slightly different operator than does $V'_T$, is also taken to vanish), we now also include the isospin-dependent interaction (coefficient $V_I$) of Eq.~(\ref{eq:FullHam}) and allow the value $m_\de$ to vary accordingly.  In this framework, the original $\delta = sq$ diquark mass $m_\delta = 0.710~\mathrm{GeV}$ used in Ref.~\cite{Jafarzade:2025qvx} produces  mass eigenvalues that lie approximately $50~\mathrm{MeV}$ below the physical masses of $\phi(2170)$ and $\eta(2370)$ when $V_I = -21.5~\mathrm{MeV}$.  An analogous effect that requires a higher diquark mass when including the isospin-dependent interaction has also been reported in the $S$-wave $c\bar{c}q\bar{q}$ tetraquark sector~\cite{Giron:2021sla}.

We offset this downward mass-eigenvalue shift that occurs for $s\bar s q\bar q$ states when spin- and isospin-dependent interactions are included by increasing the $sq$ diquark mass to $m_\de = 0.745$~GeV\@.  The masses of the \(1^{--}\) and \(0^{-+}\) states can then easily be brought into agreement with the observed values of \(\phi(2170)\) and \(\eta(2370)\), respectively.  In these fits, we take $\phi(2170)$ to be the lightest $1^{--}$ and $\eta(2370)$ to be the heavier $0^{-+}$ $s\bar s q\bar q$ eigenstate.

 To estimate the theoretical uncertainties, we vary each of the four model parameters independently by $\pm 10\%$, keeping all other parameters fixed during each variation  (Table~\ref{tab:VaryParams}).  The resulting mass  sensitivities are small for most parameters: For the $0^{-+}$ state, the dominant contributions arise from  shifts of $V_{LS}$ and $V_I$, which  change the masses by 5.8 and 6.4~MeV, respectively, while $\kappa_{ss}$ and $V_T'$ induce only
sub-MeV effects. For the $1^{--}$ state, the  sensitivities are almost entirely governed
by $V_I$, which produces a change of 19~MeV, with all other parameters contributing
$< 1$~MeV\@.   This sensitivity to $V_I$ is particularly interesting: Until isospin-partner tetraquark states [{\it e.g.}, hidden-strange analogues to $Z_c(3900)$] are identified, the precise value of $V_I$ will be difficult to establish.

\begin{table}[h!]
\centering
\caption{Mass sensitivities $\Delta M$ in (MeV) of the $0^{-+}$ and $1^{--}$ states to $\pm10\%$ variations of the model parameters. Each range corresponds to  the result of varying only the indicated parameter.}
\begin{tabular}{l@{\hskip 3em}c@{\hskip 3em}c}
\hline\hline
Parameter & $\Delta M(0^{-+})$ & $\Delta M(1^{--})$ \\
\hline
$\kappa_{ss}$     &  $  0.2$ & $  0.6$ \\
$V_{LS}$          & $  5.8$  & $  2\times 10^{-4}$  \\
$V_T'$            & $  5\times 10^{-3}$ & $  1\times 10^{-3}$ \\
$V_I$   & $  6.4$        & $  19.0$          \\
\hline\hline
\end{tabular}
\label{tab:VaryParams}
\end{table}

These states are further refined within the diabatic formalism by incorporating nearby hadronic thresholds. The results for $\phi(2170)$ and $\eta(2370)$ are presented in Tables~\ref{tab:phi} and \ref{tab:X(2370)}, respectively.  The \(\phi(2170)\) mass lies not far from the \(\Lambda\bar{\Lambda}\) threshold, which is therefore explicitly included in the calculation.  The corresponding $\phi(2170)$ wave function is again dominated (Table~\ref{tab:phi}) by the \(\delta\bar{\delta}\) component \((\approx 85\%)\), with smaller contributions from \(\Lambda\bar{\Lambda}\) \((\approx 4\%)\) and \(\eta(958) f_{1}(1285)\) \((\approx 3\%)\). Thus, the $\de\bde$ configuration remains the dominant component, even when baryon–antibaryon and di–meson channels are included.

\begin{table}[h!]
    \caption{Results for the energy eigenvalue $E$, state size $\langle r \rangle$, and diquark–antidiquark and di–meson state content (\%) of \(\phi(2170)\), assumed to be the lightest $1^{--}$ $s\bar s q\bar q$ $1P$ state [via Eq.~(\ref{eq:1--})] within the diabatic dynamical diquark model.
}
    \label{tab:phi}
    \centering
    \renewcommand{\arraystretch}{1.2}  
    \begin{tabular}{c@{\hskip 4em}r}
%    {\columnwidth}{@{\extracolsep{\fill}}lcc@{}}
       \hline\hline
$E$ (GeV) & 2.132  \\
$\langle r \rangle$ (fm)  & 0.971   \\
$\delta\bar{\delta}$ & 84.6   \\
$\Lambda\bar{\Lambda}$       &  4.0      \\
$\eta(958)f_1(1285)$    & 3.1      \\
\(\eta(958) \eta(1295)\)       & 1.2  \\
\(\omega \eta(1475)\)   & 1.0        \\
\(\omega f_2(1525)\)  & 1.6          \\
\(\phi f_1(1285)\)   & 1.5    \\
\hline\hline 
    \end{tabular}
\end{table}

For \(\eta(2370)\), the \(\delta\bar{\delta}\) fraction is nearly 89\% (Table~\ref{tab:X(2370)}), while the next two subleading contributions arise from \(\phi h_{1}(1415)\) \((\approx 5\%)\) and \(\phi  f_1(1285)\) \((\approx 4\%)\) channels. This state is also interpreted as a $0^{-+}$ $s\bar{s}q\bar{q}$ state in a recent QCD sum-rule analysis~\cite{Wang:2025nme}.

\begin{table}[h]
    \caption{Results for the energy eigenvalue $E$, state size $\langle r \rangle$, and diquark–antidiquark and di–meson state content (\%) of \(\eta(2370)\), assumed to be the heavier $0^{-+}$ $s\bar s q\bar q$ $1P$ state [via Eq.~(\ref{eq:0-+})] within the diabatic dynamical diquark model.}
    \label{tab:X(2370)}
    \centering
    \renewcommand{\arraystretch}{1.2}  
    \begin{tabular}{c@{\hskip 4em}r}
       \hline\hline
$E$ (GeV) & 2.352  \\
$\langle r \rangle$ (fm)  & 0.859   \\
$\delta\bar{\delta}$ & 88.7   \\
 \(\phi h_1(1415)\)           & 4.9\\
\(\phi f_1(1285)\)   & 3.6    \\
\hline\hline 
    \end{tabular}
\end{table}

In both cases, the adiabatic model produces a perfect fit to the measured masses, but the diabatic corrections pull them about 20--30~MeV lower.  However, this discrepancy is merely an artifact of the two-stage fitting process; a global diabatic fit could easily maintain the perfect fit. Indeed, a fully self-consistent global diabatic fit treats  all states in each multiplet simultaneously and includes  all di-hadron channel couplings from the start, providing a consistent description of masses and wave functions across the spectrum. In practice, the two-stage procedure  (first adiabatic, then diabatic) captures the main effects of mixing and threshold coupling, while producing only modest mass shifts.   This claim is supported by analysis in the hidden-charm sector, where the diabatic approach has been extended to treat the tetraquark states as poles in di-hadron scattering amplitudes~\cite{Lebed:2023kbm}, and yet the resonant peak positions are nevertheless found to shift rather modestly compared to either the adiabatic results or to diabatic results in which mass shifts (and decay widths) are computed using conventional Fano-type~\cite{Fano:1961zz} coupled-channel methods~\cite{Lebed:2024rsi}. Even with a global fit adjusted to the physical masses of $\phi(2170)$ and $\eta(2370)$, the overall structure of these states is not expected to change significantly.  The compact $\de\bde$ configuration would continue to dominate, with the fit mainly refining numerical details.

 Using the same parameter set as presented in  this subsection and assuming no mixing, we predict additional isoscalar $1^{--}$  $s\bar s q\bar q$ tetraquark states with masses of 2397~MeV, 2377~MeV, and 2340~MeV, as well as a $0^{-+}$ state at 2088~MeV\@.  We also predict an exotic ($0^{--}$) tetraquark state with a mass of 2359~MeV\@. Nevertheless, a careful analysis of the known PDG isoscalar states with the same quantum numbers (interpreted as tetraquark candidates) is needed in order to constrain the model parameters directly, rather than  by relying upon comparison with another theoretical approach  (such as QCD sum rules),  especially for  the fully strange tetraquarks  to be discussed next.

\subsection{$S$-Wave \(s\bar{s}s\bar{s}\) States} \label{sec:Swavessss}

In the \textit{$S$-wave fully strange} sector, we predict states with \(J^{PC}=0^{++}\) and \(2^{++}\) at \(2.19\)~GeV and \(2.18\)~GeV, respectively (Table~\ref{tab:$S$-wave-ss}).  Both are dominated by the \(\delta\bar{\delta}\) component (\(\approx 90\%\)).  For the \(0^{++}\) state, the next largest contribution is from \(\eta(958)\eta(1295)\) (\(\approx 5\%\)), while for the \(2^{++}\) state this channel contributes only \( \approx 1\%\).  The suppression of the latter value arises from the angular momentum barrier: According to Table~ \ref{tab:meson_thresholds-mixed}, \(0^{++}\) $\eta \, \eta$ states couple through the \(\ell=0\) partial wave, whereas \(2^{++}\) {$\eta \, \eta$} states only appears through the \(\ell=2\) partial wave.

\begin{table}[h!]
\caption{Diabatic formalism results for $S$-wave $s\bar{s}s\bar{s}$ states: energy eigenvalues $E$, state sizes $\left< r \right>$, and diquark-antidiquark and di-meson state contents (\%).}
%; “–” indicates forbidden quantum numbers.}
\label{tab:$S$-wave-ss}
\centering
 \renewcommand{\arraystretch}{1.2}
\begin{tabular}{c@{\hskip 4em}r@{\hskip 4em}r}
\hline\hline
$J^{PC^{\vphantom{\dagger}}}$ & $0^{++}$   & $2^{++}$   \\
\hline
$E$ (GeV) & 2.188   & 2.179   \\
 
$\langle r \rangle$ (fm) & 0.545  & 0.543  \\
 
$\delta\bar{\delta}$ & 89.3 & 91.4   \\
 
$\Lambda\bar{\Lambda}$ & 1.5  & 1.5 \\
$\eta(958)f_1(1285)$    & 1.1     & 1.3   \\
$\eta(958)\eta(1295)$   & 4.5 &1.1 \\
\hline\hline
\end{tabular}
\end{table}

Experimentally, the PDG lists nearby candidates \(f_{0}(2200)\) and \(f_{2}(2150)\), both with reported widths of \(\gtrsim 200\)~MeV\@.  The same holds for the somewhat more distant $f_0(2100)$.  \(f_{0}(2200)\) has also been suggested as an $S$-wave tetraquark in Ref.~\cite{Liu:2020lpw}.

\subsection{$\bm{X}$(2300) and $\bm{f_{2}(2340)}$} \label{sec:X2300}

We also investigate the recently observed \(J^{PC}=1^{+-}\) \(X(2300)\)~\cite{BESIII:2025prl}, which has been interpreted as either a  fully strange tetraquark \cite{Wan:2025xhf} or a molecular system~\cite{Cao:2025dze}.

 Several recent studies have proposed alternative interpretations for the newly observed $X(2300)$. The PACIAE study \cite{Cao:2025dze} simulates $e^+e^-$ collisions to generate a full partonic and 
hadronic final state, including parton rescattering, hadronization, and hadronic rescattering. Using a phase-space coalescence approach, hypothetical candidates for 
$X(2300)$ are assembled under different  possible internal structures: an excited $s\bar s$, a compact $ss\bar s\bar{s}$ tetraquark, and a hadro-strangeonium ($\phi\eta$ or $\phi\eta'$) state. For each configuration, the model computes production yields and kinematic distributions (rapidity and $p_T$), finding that the resulting yields and spectra differ significantly among the three hypotheses: The excited $s\bar s$ and compact tetraquark configurations are produced at observable rates, whereas the hadro-strangeonium candidate is strongly suppressed.  Work using the modified Godfrey--Isgur (MGI) approach \cite{Hao:2024nvx} constructs a relativized $s\bar{s}$ 
spectrum with a screened linear potential to mimic vacuum-polarization effects, which lowers the predicted masses of highly excited states  below naive quark-model expectations. Using the $^{3}P_0$ quark-pair creation model,  Ref.~\cite{Hao:2024nvx} then calculates OZI-allowed two-body decay widths, assigning $X(2300)$ as the third radial excitation of the $P_1$ axial-vector state 
($3^1P_1$), and predicts $K\bar K^*$, $K^*\bar K^*$, 
$K^*\bar K_{1B}$, and $K\bar K_2^*(1430)$  as the dominant decay modes.

To test its structure, we vary $m_\de$ from $0.95$~GeV~(its value in the adiabatic calculation~\cite{Jafarzade:2025qvx}) to $1.0$~GeV, in order to reproduce the observed $X(2300)$ and include nearby closed thresholds.  Our results (Table~\ref{tab:X(2300)}) show that even after threshold effects are included, \(X(2300)\) remains dominated by its \(\de\bde\) component (\(\approx 83\%\)), followed by two \(\eta\phi\) channels with excited $\eta$'s (each \( \approx 7\%\)).  These results, analogous to those in previous cases, support the interpretation of $X(2300)$ as a compact \(S\)-wave \(s\bar{s}s\bar{s}\), $J^{PC} = 1^{+-}$ state.

Furthermore, Table~\ref{tab:X(2300)} also shows that the shift to $m_\de = 1.0$~GeV nicely accommodates \(f_{2}(2340)\) ($J^{PC} = 2^{++}$), which is also found to be an $S$-wave $s\bar s s\bar s$ state that is predominantly \(\delta\bar{\delta}\) \((\approx 93\%)\), with subleading contributions from \(\eta(958)h_{1}(1415)\) and \(\eta(958)f_{1}(1420)\) (each \(\gtrsim1\%)\).

\begin{table}[h]
    \caption{Results for the energy eigenvalues $E$, state sizes $\left< r \right>$, and state contents of $X(2300)$  and $f_2(2340)$ within the diabatic dynamical diquark model with $m_\delta=1.0$~GeV\@.  Entries $< 1\%$ are left blank.}
    %“–” indicates forbidden quantum numbers.}
    \label{tab:X(2300)}
    \centering
    \renewcommand{\arraystretch}{1.2}  
    \begin{tabular}{c@{\hskip 4em}r@{\hskip 4em}r}
       \hline\hline
       $J^{PC^{\vphantom{\dagger}}}$&$1^{+-}$ &$2^{++}$\\ \hline
$E$ (GeV) & 2.271& 2.307 \\
$\langle r \rangle$ (fm)  & 0.567  &0.517 \\
$\delta\bar{\delta}$ & 82.7& 92.8 \\
 \(\eta(1295)\phi \) & 6.7& \\%\text{not included} \\
  $\eta(958)\,h_1(1415)$   &  & 1.3\\
\(\eta(958)f_1(1420)\)   &     & 1.1    \\

\( \eta(1475)\phi\)            & 7.1 &      \\

\hline\hline 
    \end{tabular}
\end{table}

Notably, among the tensor states with masses around 2.3 GeV observed also in the channel \(J/\psi \to \gamma\phi\phi\) \cite{BESIII:2016qzq}, the significantly larger decay width of $f_{2}(2340)$ compared with that of $f_{2}(2300)$ further supports  an exotic  interpretation. The exotic nature of $f_{2}(2340)$ has also been recently examined in the context of a tensor glueball, but analyses of its decay channels within the chiral model disfavor this  assignment~\cite{Vereijken:2023jor}.

\subsection{$P$-Wave \(s\bar{s}s\bar{s}\) States}

Finally, in the \textit{$P$-wave fully strange} sector, we take the \(\Xi \, \bar{\Xi}\) threshold as the lightest one (Table~\ref{tab:ss-Meson thresholds}), because the following mass relation  for the $0^{-+}$ state when the spin-dependent terms are included~\cite{Jafarzade:2025qvx},
\begin{align}
    M_{0^{-+}}=M_0(1P)+\kappa_{ss}-2V_{LS} - 8V'_T \,,
\end{align}
with $V_{LS} > 0$  and $V'_T$ assumed consistent with 0 as in Sec.~\ref{sec:phi}, suggests that this state is among the lightest in the multiplet.
Using the parameter values introduced in Sec.~\ref{sec:phi}, the pseudoscalar tetraquark is predicted to have a mass of 2.62~GeV within the adiabatic approximation.  In the diabatic formalism, the spectrum remains strongly dominated by the $\de\bde$ configuration (Table~\ref{tab:$P$-wave ss}).  In particular, the predicted mass of the \(0^{-+}\) state agrees with the observed \(X(2600)\).  It is described primarily as a $\de\bde$ configuration ($\approx 80$\%), with the next main contribution arising from  the $\Xi \, \bar \Xi$ component ($\approx 15$\%).  Indeed, this contribution is the largest single di-hadron component we have observed in the hidden-strange sector.

Several theoretical studies have interpreted the $X(2500)$ resonance as a fully strange tetraquark ($s s \bar{s} \bar{s}$). QCD sum-rule analyses~\cite{Su:2022eun,Dong:2020okt} and quark-model calculations~\cite{Liu:2020lpw} assign it to a $P$-wave configuration with quantum numbers $J^{PC}=0^{-+}$, predicting masses in the range $2.45$--$2.55~\text{GeV}$, consistent with the BESIII measurement. These works suggest that $X(2500)$ represents the lowest $1P$ fully strange tetraquark excitation, rather than a conventional $s\bar{s}$ meson.  In our framework,  however, we reach a different conclusion: The corresponding $s \bar s s\bar s$ tetraquark configuration appears around $2.6~\text{GeV}$, supporting the interpretation of the nearby $X(2600)$ state as the likelier $s\bar s s\bar s$ state, especially since the quantum numbers $J^{PC} = 0^{-+}, 2^{-+}$, both members of the $1P$ $s\bar s s\bar s$ multiplet, are favored~\cite{BESIII:2022sfx}.

The $1^{--}$ sector is handled analogously and gives very similar results (Table~\ref{tab:$P$-wave ss}); in this case, however, we note that the $s\bar s s\bar s$ sector contains two $1^{--}$ states (see Sec.~\ref{sec:States}), and so we input the average of their mass eigenvalues obtained from the adiabatic calculation (2.62~GeV) into our diabatic calculation.  The resulting value of 2.584~GeV in Table~\ref{tab:$P$-wave ss} provides a prediction for the mass average of the two $1^{--}$ $P$-wave $s\bar s s\bar s$ states in the diabatic model.

\begin{table}[h!]
\caption{Diabatic formalism results for $P$-wave $s\bar{s}s\bar{s}$ states: energy eigenvalues $E$, state sizes $\left< r \right>$, and diquark-antidiquark and di-meson state contents (\%).  “–” indicates forbidden quantum numbers.}
\label{tab:$P$-wave ss}
    \renewcommand{\arraystretch}{1.2}  
\begin{tabular}{c@{\hskip 4em}r@{\hskip 4em}r}
\hline\hline
$J^{PC^{\vphantom{\dagger}}}$ & $1^{--}$ & $0^{-+}$\\ \hline
$E$ (GeV) & 2.584 &2.602\\
$\langle r \rangle$ (fm)  & 0.855& 0.855\\
$\delta\bar{\delta}$   & 79.5 & 80.0 \\
$\Xi \, \bar \Xi$               & 10.4 & 15.2 \\
$f_1(1285)h_1(1415)$        & 1.7 & 1.6\\
$\eta(1295)h_1(1415)$    & 2.9 & $-$ \\
$f_1(1285)f_1(1420)$     & 1.4 & 1.3 \\ 
$\eta(1295)f_1(1420)$     &2.3   & $-$ \\
\hline\hline
\end{tabular}
\end{table}

\section{Conclusions}
\label{sec:Concl}

In this work, we have studied the spectroscopy and internal structure of hidden-strangeness tetraquarks within the diabatic dynamical diquark framework, taking into account the effects of nearby hadronic thresholds that can contribute to the structure of the states.  Our analysis shows that both $S$- and $P$-wave tetraquarks are predominantly composed of compact diquark-antidiquark configurations, with numerically smaller (typically $< 10 \%$) contributions arising from di-meson and baryon-antibaryon channels.  Threshold effects generally reduce the mass eigenvalues predicted from the adiabatic approach by a few tens of MeV, but their effects become negligible when the tetraquark lies sufficiently far above the nearest meson threshold.

For the $S$-wave \(s\bar{s}q\bar{q}\) and \(s\bar{s}s\bar{s}\) states, the \(\delta\bar{\delta}\) component dominates, and the small  state size $\left< r \right>$ values confirm a compact tetraquark interpretation.  In particular, \(f_2(2340)\) and \(X(2300)\) are found to retain predominantly diquark-antidiquark character, even after including threshold effects. 

In the $P$-wave sector, \(\phi(2170)\) and \(\eta(2370)\) are found to be consistent with compact $s \bar s q\bar q$ tetraquark structures.   We also find that $X(2600)$ fits well as a $P$-wave $s\bar s s\bar s$ state with quantum numbers $0^{-+}$.  Of all the states considered in this work, the largest mixing observed is the $\Xi$-$\bar \Xi$ content of $X$(2600), which is found to be almost 15\%.

To this point, the only distinction between members of dynamical-diquark-model multiplets (such as $0^{++}$ and $1^{++}$ in the $S$-wave, or $0^{-+}$ and $1^{--}$ in the $P$-wave) implemented uniformly in the hidden-strangeness sector appears through the use of distinct masses for the component hadrons in the di-hadron thresholds.  A full analysis of the multiplet spectral structures ({\it e.g.}, studies of $P$-wave states with unusual quantum numbers, such as $0^{--}$,  which have been ignored here) requires a complete treatment of fine-structure effects, as has been carried out in the heavy-quark sector~\cite{Giron:2019cfc,Giron:2020fvd,Giron:2020qpb,Giron:2020wpx,Giron:2021sla}.  Some of the necessary steps have been presented here, such as the use of Eqs.~(\ref{eq:1--})--(\ref{eq:0-+})  for the special cases of $1^{--}$ and $0^{-+}$ $P$-wave $s\bar s q\bar q$ states, but a full  analysis remains to be completed.

Overall, our results suggest that exotic tetraquark configurations may  be abundant not only in the heavy-quark sector, but also among hidden-strangeness states.  However, distinct and well-separated thresholds do not appear to produce the same impact for such states as they do in the heavy-quark sector.  The diabatic dynamical diquark model provides a unified framework for interpreting these resonances, yielding predictions for masses,  state compositions, and spatial structures that can be confronted with ongoing measurements  from BESIII and Jefferson Lab, as well as  at future facilities such as the Electron–Ion Collider.

\begin{acknowledgments}
SJ thanks Steven Martinez for discussions and for sharing his numerical results on the topic.  RFL acknowledges support by the National Science Foundation (NSF) under Grant No.\ PHY-2405262.  SJ acknowledges support by the U.S.\ Department of Energy ExoHad Topical Collaboration under Grant No.\ DE-SC0023598.  This work contributes to the goals of the ExoHad Collaboration.
\end{acknowledgments}

\bibliographystyle{apsrev4-2}
\bibliography{main}
 
\end{document}